\documentclass[10pt,conference,a4paper]{IEEEtran}
\UseRawInputEncoding

\usepackage{cite}
\usepackage{amsmath,amssymb,amsfonts}

\usepackage{algorithmic}
\usepackage{graphicx}
\usepackage{textcomp}
\usepackage{xcolor}

\usepackage{float}
\usepackage{caption}
\usepackage{subcaption}
\usepackage{url}

\usepackage[boxed,ruled,vlined,linesnumbered,commentsnumbered, noresetcount]{algorithm2e}

\usepackage{soul}

\usepackage{tabularx}
\usepackage{ragged2e} 
\newcolumntype{L}{>{\RaggedRight}X} 

\usepackage{comment}
\usepackage{makecell} 
\usepackage{multirow}

\begin{document}

\title{Scalability Analysis of 5G-TSN Applications in Indoor Factory Settings}

\author{Kouros Zanbouri, Md. Noor-A-Rahim, Dirk Pesch\\
School of Computer Science and Information Technology\\ University College Cork, Ireland
}

\maketitle

\begin{abstract}
While technologies such as Time-Sensitive Networking (TSN) improve deterministic behaviour, real-time functionality, and robustness of Ethernet, future industrial networks aim to be increasingly wireless. While wireless networks facilitate mobility, reduce cost, and simplify deployment, they do not always provide stringent latency constraints and highly dependable data transmission as required by many manufacturing systems. The advent of 5G, with its Ultra-Reliable Low-Latency Communication (URLLC) capabilities, offers potential for wireless industrial networks. 5G offers elevated data throughput, very low latency, and negligible jitter. As 5G networks typically include wired connections from the base station to the core network, integration of 5G with time-sensitive networking is essential to provide rigorous QoS standards. This paper assesses the scalability of 5G-TSN for various indoor factory applications and conditions using OMNET++ simulation. Our research shows that 5G-TSN has the potential to provide bounded delay for latency-sensitive applications in scalable indoor factory settings.
\end{abstract}



\section{Introduction}
With the rise of the Industrial Internet of Things (IIoT) and the push towards Industry 4.0, efforts are being made to enhance the flexibility, interoperability, and integration of industrial automation systems. The future of smart manufacturing will see the emergence of dynamic, reconfigurable production systems capable of rapid adaptation to evolving industry requirements. These environments will leverage technologies such as mobile sensing devices, robotic arms, automated guided vehicles (AGVs), and drones to facilitate the customization of products. Since traditional wired automation systems lack the adaptability needed for such changes, wireless technologies are becoming crucial to ensure real-time, deterministic connectivity for industrial processes.

Wireless technologies are advancing to support the demanding needs of industrial environments. Wireless TSN, building on the success of wired TSN, is emerging as a promising solution \cite{10735349}. 5G and 802.11 are potential candidates for enabling wireless TSN. However, 5G, with its ultra-reliable, low-latency communication capabilities specifically designed for real-time communication through its URLLC profile, appears to be the current front-runner to become the wireless technology for Industry 4.0 applications.

5G and TSN integration promises to significantly advance real-time industrial communication. However, challenges such as precise time synchronization and efficient resource allocation remain. While early testbeds to study 5G-TSN performance are still under construction, further research is necessary to assess the scalability of 5G-TSN in industrial settings.
While there are several studies on 5G-TSN systems, to the best of our knowledge, this is the first study to examine the scalability of 3GPP standardised 5G-TSN in a wireless manufacturing environment. 
The conducted simulation-based study provides a performance assessment of an industrial use case by integrating an indoor manufacturing model into a 5G-TSN industrial network model transmitting TSN traffic over 5G. Our research presents the following key contributions: 

\begin{itemize}
    \item Integration of the 3GPP TR 38.901 Indoor Factory Profile into a 5G-TSN network model to simulate a realistic wireless indoor factory environment.

    \item In-depth performance analysis across various use cases, including both general and specific scenarios, in different factory environments.

    \item Examination of the scalability and impact of various Indoor Factory profiles, assessing how factors such as scaling up, high device density, and extensive data throughput, affect network performance.
\end{itemize}

The remainder of this paper is organized as follows. In Section \ref{sec:Preliminaries}, we discuss the fundamental concepts of TSN and 5G. In Section \ref{sec:Model&Scenarios}, we describe the models of specific indoor factory environments, along with models of selected use cases of 5G-TSN. Section \ref{sec:Simulation} details the experimental results and performance evaluation. Finally, Section \ref{sec:Conclusion} summarizes the key findings and discusses potential future research directions.

\section{Preliminaries} \label{sec:Preliminaries}

\subsection{Time-Sensitive Networking}
The IEEE 802.1 TSN standards enhance standard Ethernet functionalities by including characteristics that guarantee determinism, dependability, and effective data handling. The fundamental components of TSN are time synchronisation, high availability, ultra-reliability, bounded low latency, and resource management as discussed in the following.
\subsubsection{Time-synchronization}
The IEEE~802.1AS standard defines a protocol for synchronizing distributed nodes in an IEEE 802 network, using a generalized Precision Time Protocol (gPTP) profile based on IEEE 1588v2. A best master clock algorithm identifies the most accurate clock, which propagates time information to all supporting nodes. TSN synchronization requirements are driven by network scheduling and application demands, with strict synchronization ensuring guaranteed maximum latency for time-critical traffic.
\subsubsection{High availability and ultra-reliability}
To achieve high availability and ultra-reliability, various standards and mechanisms, such as IEEE~802.1Qca, IEEE~802.1Qci and IEEE~802.1CB Frame Replication and Elimination for Reliability (FRER), can be employed. For instance, using FRER, source end stations may transmit duplicated frames (pertaining to a given stream) across multiple paths in the network to ensure seamless redundancy and enhance the reliability of real-time communications.
\subsubsection{Bounded low latency}
TSN-based networks ensure bounded low latency and zero congestion loss, enabling the prioritization of time-sensitive packets during transmission across the network. To accomplish this, a range of TSN mechanisms including traffic scheduling, traffic shaping, and frame pre-emption have been developed within the IEEE 802.1 standards framework, encompassing specifications such as 802.1Qav, 802.1Qbv, 802.1Qbu, 802.1Qch, and 802.1Qcr.

\subsubsection{Resource management}
Achieving TSN's low latency and reliability requires proper configuration of network elements and efficient management of bandwidth, paths, and schedules. The IEEE 802.1Q standard defines traffic management methods for bridged Ethernet networks. IEEE 802.1Qcc enhances TSN network setup and management, while the IEEE 802.1Qat Stream Reservation Protocol (SRP) reserves bandwidth and schedules resources for time-sensitive applications using MMRP, MVRP, and MSRP. Complementing these, IEEE 802.1Qca focuses on path control and bandwidth reservation for TSN flows, with IEEE 802.1Qcc improving SRP functionalities.

\subsection{5G and wireless TSN}
Currently, 5G is recognized as the leading candidate for implementing wireless TSN solutions. In its Release 16, the 3rd Generation Partnership Project (3GPP) laid the groundwork for enabling mobile and wireless communications to achieve the speed and reliability required by Industry 4.0. This release introduces mechanisms to deliver ultra-reliable low-latency communication (URLLC) with sub-1~ms latency and highly precise synchronization, with an accuracy better than 1 \(\mu\)s, leveraging IEEE 802.1AS. Integration with IEEE~802.1 TSN is facilitated through the 5G Bridge framework, which incorporates critical elements such as the device-side TSN Translator (DS-TT), user equipment (UE), the radio access network (RAN), and the core network (5GC). The core network architecture is divided into user plane functions (UPFs), featuring the network-side TSN Translator (NW-TT), and control plane functions that include components for managing access, mobility (AMF), sessions (SMF), policies (PCF), and unified data (UDM) \cite{9855453}. It is essential to understand the influence of wireless channels when incorporating TSN into 5G networks to ensure the effective management of TSN performance requirements. The next section delves into the unique characteristics of indoor factory environments and explores the use-case of 5G-TSN within this context.

\section{System Model and Scenarios} \label{sec:Model&Scenarios}
\subsection{Indoor Factory Channel Modelling}
\begin{figure}
    \centering
    \includegraphics[width=1\linewidth]{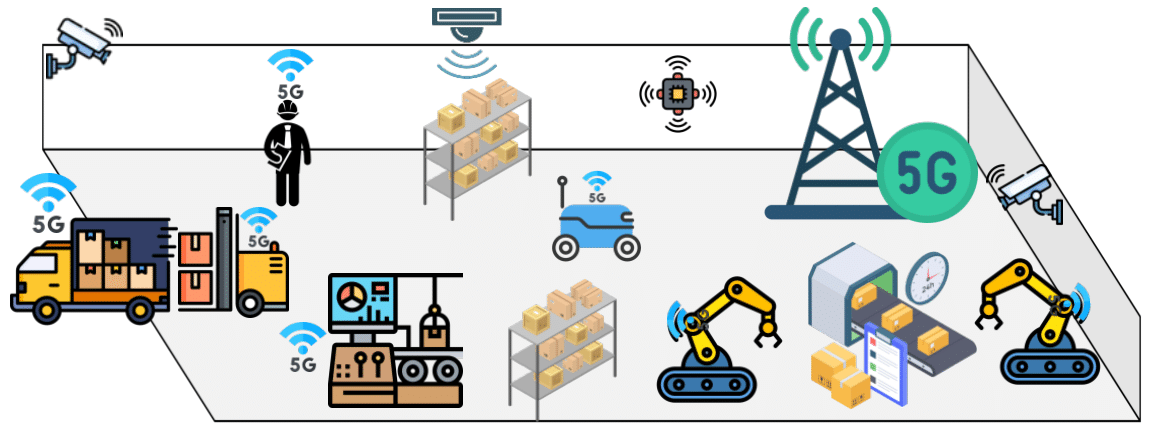}
    \caption{5G-TSN Indoor Factory Setting}
    \label{fig:5G-TSN-Overview}
\end{figure}

Since wireless channels are highly influenced by their surroundings, it is essential to consider the critical environmental features of the intended deployment area when modelling a wireless channel for that specific setting. 3GPP has outlined comprehensive channel models for the 5G New Radio (NR) system across various environments in its report, 3GPP TR 38.901. \cite{3gpp.38.901}. The report includes indoor factory scenarios, detailing factory buildings of varying sizes and densities, featuring different levels of clutter such as machinery, assembly lines, and storage shelves. As outlined in the technical report, several indoor factory (InF) profiles are defined: InF-SL (sparse clutter, low base station), InF-DL (dense clutter, low base station), InF-SH (sparse clutter, high base station), InF-DH (dense clutter, high base station), and InF-HH (high transmitter, high receiver).
Fig. \ref{fig:5G-TSN-Overview} shows a 5G-TSN network in an industrial environment illustrating the elements of the indoor factory profile with both LOS and NLOS scenarios.

The path loss and line-of-sight (LOS) probability models for the 3GPP TR 38.901 channel models consider factors like antenna height and the density of objects in the environment. While the indoor office scenarios typically involve objects like people, light partitions, desks, and chairs with limited heights, the InF scenario features larger machinery that can significantly block signals. In the InF, the ceiling height also varies widely, from 5 to 25 meters. To account for these differences, the 3GPP report uses terms such as ``clutter-embedded'' and ``clutter-elevated'' to describe the height of the user terminal (UT) and base station (BS). For path loss, the model considers one LOS and four NLOS (Non-line-of-sight ) scenarios to reflect the different blocking conditions that can occur in industrial environments. The LOS Path loss of InF is calculated using the following Eq. \ref{eq:(1)}:
\begin{equation}
\label{eq:(1)}
\begin{aligned}
PL_{\text{LOS}} = 31.84 + 21.5 \cdot \log_{10} \left({d_{3D}} \right) + 19 \log_{10} \left({f_c} \right) \\ 
\sigma_{SF} = 4
\end{aligned}
\end{equation}
where \(d_{3D}\) is the 3D distance between transmitter and receiver which is in meters and is constrained by \(1 \leq d_{3D} \leq 600\) meters, \(f_c\) denotes carrier frequency which is in GHz. In addition, shadow fading standard deviation (in dB) is represented by \(\sigma_{SF}\).
NLOS path loss of the InF-SL is evaluated using Eq. \ref{eq:InF-SL}:
\begin{equation}
\label{eq:InF-SL}
\begin{aligned}
PL &= 33 + 25.5 \log_{10} \left({d_{3D}} \right) + 20 \log_{10} \left({f_c} \right) \\
PL_{\text{NLOS}} &= \text{max}\, (PL, PL_{LOS}), \sigma_{SF} = 5.7
\end{aligned}
\end{equation}
For InF-DL, the NLOS path loss is calculated using Eq. \ref{eq:InF-DL}:
\begin{equation}
\label{eq:InF-DL}
\begin{aligned}
PL &= 18.6 + 35.7 \log_{10}(d_{3D}) + 20 \log_{10}(f_{c}) \\
PL_{\text{NLOS}} &= \text{max}\, (PL, PL_{\text{LOS}}, PL_{\text{InF-SL}}), \sigma_{SF} = 7.2
\end{aligned}
\end{equation}
InF-SH, on the other hand, has a different path loss calculation for NLOS, which uses the following Eq. \ref{eq:InF-SH}:
\begin{equation}
\label{eq:InF-SH}
\begin{aligned}
PL &= 32.4 + 23.0 \log_{10}(d_{3D}) + 20 \log_{10}(f_{c}) \\
PL_{\text{NLOS}} &= \text{max}\, (PL, PL_{\text{LOS}}) , \sigma_{SF} = 5.9
\end{aligned}
\end{equation}
Finally, for InF-DH, the following Eq. \ref{eq:InF-DH} is considered to evaluate NLOS:
\begin{equation}
\label{eq:InF-DH}
\begin{aligned}
PL &= 33.63 + 21.9 \log_{10}(d_{3D}) + 20 \log_{10}(f_{c}) \\
PL_{\text{NLOS}} &= \text{max}\, (PL, PL_{\text{LOS}}), \sigma_{SF} = 4
\end{aligned}
\end{equation}
The LOS probability for InF-SL, InF-SH, InF-DL, and InF-DH profiles can be calculated using Eq. \ref{eq:los_prob}, \ref{eq:k_subsec}, \ref{eq:los_hh}:
\begin{equation}
\text{Pr}_{\text{LOS, subsec}}(d_{2D}) = \exp\left(-\frac{d_{2D}}{k_{\text{subsec}}}\right)
\label{eq:los_prob}
\end{equation}
where \begin{equation}
k_{\text{subsec}} = 
\begin{cases} 
    \frac{-d_{\text{clutter}}}{\ln(1 - r)} & \text{for InF-SL and InF-DL} \\
    \frac{-d_{\text{clutter}}}{\ln(1 - r)} \cdot \frac{h_{\text{BS}} - h_{\text{UT}}}{h_{c} - h_{\text{UT}}} & \text{for InF-SH and InF-DH}
\end{cases}
\label{eq:k_subsec}
\end{equation}

The parameters $d_{\text{clutter}}$, $r$, and $h_c$ are specified in Table 7.2-4 of 3GPP TR 38.901 \cite{3gpp.38.901}. \(h_{\text{BS}}\) and \(h_{\text{UT}}\) represent the antenna heights for the base station and user terminal, respectively. For InF-HH, the LOS probability is \begin{equation} \text{Pr}_{\text{LOS}} = 1 \label{eq:los_hh} \end{equation}

\subsection{Scenario/Use-Case}

\begin{figure}
    \centering
    \includegraphics[width=1\linewidth]{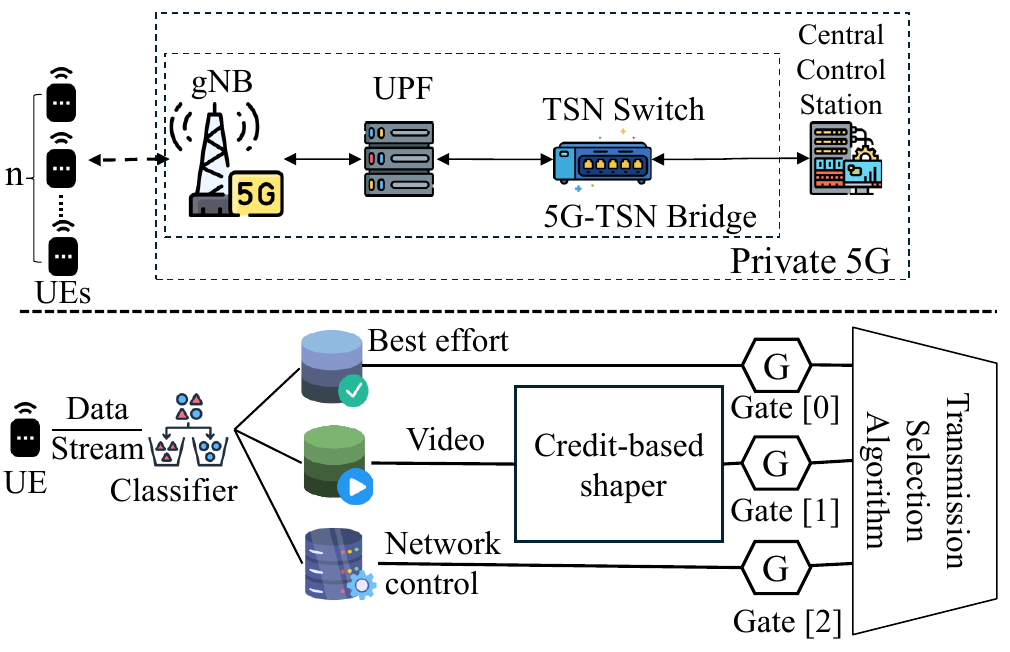}
    \caption{5G-TSN network topology with n UEs connected to the TSN domain, with a focus on traffic transmission at the TSN Switch}
    \label{fig:5G-TSN-architecture}
\end{figure}

AGVs, originally developed for material handling in factories, have evolved into key mobile devices for flexible manufacturing. With 5G connectivity, they enhance operational efficiency by autonomously inspecting factories, improving safety, and streamlining logistics. Deploying multiple AGVs indoors optimizes workflow, making them a valuable investment for companies like Amazon, DHL, and Tesla. While real-world applications exist, researchers continue exploring scalability and implementation challenges for 5G-powered AGVs and AMRs. A notable example is a 5G-connected mobile industrial robot (MiR) equipped with a battery, LiDAR camera, video processing unit, and private 5G network access \cite{9247159}. The setup integrates mobile edge computing for remote access and enhanced functionality, as shown in Fig.~\ref{fig:5G-TSN-architecture}.

In our 5G-TSN network scenario, traffic is categorized into three types. Automation Network Control (NC), Video - to partly support navigation, and Best Effort (BE). NC and Video are TSN traffic streams, with NC assigned the highest priority, Video a medium priority, and BE, a non-TSN stream, the lowest priority. NC traffic is periodic, where strict real-time behaviour is required. This type of traffic carries time-sensitive data crucial to the AGV's navigation, control, and safety, including information like position, speed, obstacle detection, and commands from the central control system. Video traffic is periodic and is streamed from the AGV's camera to support tasks such as object recognition, visual inspection, and remote monitoring. The consistent periodicity of this traffic stream ensures stable video quality and minimizes latency. BE handles non-critical data, such as telemetry, diagnostics, or software updates, which are not essential for immediate AGV operations. This traffic often alternates between inactivity and bursts of high activity, making it suitable for low-priority background tasks where timing is flexible.

To accurately model the behaviour of applications in our network, randomness is introduced in the initial production offset and start time of each traffic stream. The initial production offset determines when an application begins transmitting data after initialization, allowing for variability that simulates real-world delays in data generation. For instance, NC traffic, which requires strict real-time behaviour, has a smaller offset, enabling it to start transmission almost immediately. In contrast, Video and BE traffic experience longer offsets, with BE assigned the longest delay before transmission begins. Additionally, the start time, which dictates when the data transmission actually starts, varies for each application. NC traffic typically begins without delay, Video starts after a short delay, and BE starts with a more randomized time. This randomness in both the initial production offset and start time is crucial for ensuring that each traffic stream behaves realistically, reflecting the unpredictable nature of network traffic in real-world environments, where synchronization is rare, and application timing varies based on priority and type.

Incoming traffic is sorted utilizing Strict Priority Queuing (IEEE 802.1Q) into different priority levels based on the 3-bit Priority Code Point (PCP) field in the VLAN tag. This approach ensures that traffic marked with the highest priority is transmitted first, while lower-priority traffic is queued until higher-priority queues are cleared. This method assigns the highest priority to NC traffic to guarantee minimal queuing delays and ensures uninterrupted transmission even during high network activity. The video traffic is managed using a queue with a Credit Based Shaper (CBS) mechanism. While the queue is inactive, it gains credit at a rate defined by an idleSlope parameter. When sufficient credit is accumulated and a chance to transmit occurs, the queue uses this credit to send data at the rate specified by sendSlope. The idleSlope and sendSlope are calculated using the following Eq. \ref{eq:idleSlope}, \ref{eq:sendSlope}: \begin{equation} \text{idleSlope} = \frac{8 \times  \text{Max Packet Length}\times \text{Data Rate}} {1000} \label{eq:idleSlope} \end{equation}
\begin{equation} \text{sendSlope} = \text{idleSlope} - \text{bitrate}\label{eq:sendSlope} \end{equation}
This classification of traffic enables efficient allocation of network resources, ensuring that NC and Video streams receive priority, while BE traffic is managed opportunistically to optimize overall network performance in time-sensitive environments.

\section{Simulation Results and Discussion} \label{sec:Simulation}
For our simulation, we employed the OMNET++ 6.0.3 simulation environment \cite{10.4108/ICST.SIMUTOOLS2008.3027}, incorporating the iNet 4.5.3 \cite{Mészáros2019}, Simu5G 1.2.2 \cite{Simu5G}, and 5GTQ \cite{10333533} frameworks. The specific parameters employed in our simulation are detailed in Table~\ref{tab:simulation}.
\begin{table}
    \centering
    \begin{tabular}{|c|c|}
    \hline
       \textbf{Parameter} & \textbf{Value} \\ \hline
        No. of UEs & 5, 10, 25, 50\\ \hline
        No. of gNBs & 1 \\ \hline
        gNB Tx power & 23 dBm \\ \hline
        Target Bler  & 0.01 \\ \hline
        Physical environment & FlatGround \\ \hline
        Carrier frequency & 5.9 GHz \\ \hline
        Numerology index & 4 \\ \hline
        Channel model & Indoor Factory\\ \hline
        InF profiles & InF-SL, InF-DL, InF-SH, InF-DH\\ \hline
        Mobility model & Random Waypoint Mobility\\ \hline
        Mobility speed & 0.2-1.5 mps \\ \hline
    \end{tabular} 
    \caption{5G Configuration settings for the simulated environment}
    \label{tab:simulation}
\end{table}
In the simulated 5G-TSN network, each data flow is assigned a specific Quality of Service (QoS) profile. This profile specifies the required resources, priority level, acceptable delay, maximum error rate, and other key parameters. Each QoS profile is identified by a unique 5G QoS Identifier (5QI). These 5QI values are categorized into three main groups based on their resource requirements: guaranteed bit rate (GBR), non-guaranteed bit rate (Non-GBR), and delay-critical guaranteed bit rate (DC-GBR). To prioritize various types of traffic, TSN leverages the Priority Code Point (PCP) field. In this study, PCP values range from 0 (lowest priority) to 7 (highest priority). The categorization of 5QI numbers and their associated QoS characteristics follows the standardized mapping outlined in Table 5.7.4-1 of 3GPP TS 23.501 version 18.7.0 Release 18 \cite{3gpp_ts_23_501}.

To assess the applicability of 5G-TSN in industrial scenarios, the InF profiles were implemented within OMNeT++, and the distance between the UE and the gNB was categorized into three distinct regions: d1, d2, and d3. These regions represent increasing radial distances from the gNB, measured in meters, and are aligned with typical industrial use cases. The d1 region, spanning a radius of up to 85 meters, represents environments such as standard indoor manufacturing facilities characterized by a high density of machinery, where ultra-reliable and low-latency communication is paramount. The d2 region, extending up to 170 meters, corresponds to medium-sized warehouses or large indoor manufacturing settings with moderate device density. Lastly, the d3 region, covering distances up to 255 meters, reflects large-scale industrial environments, such as expansive Amazon-style warehouses, where extensive coverage and robust device connectivity are essential. 
The delineation of these distance segments facilitates a structured evaluation of performance metrics, capturing the effects of signal strength, network performance variations, and increasing scalability with greater distance from the gNB. By dividing the use-case scenario into these three distinct distance ranges, this approach enables a more comprehensive analysis of the influence of distance on communication reliability, efficiency, and scalability within an indoor industrial environment.

Furthermore, Table \ref{tab:app_characterization} presents the application characteristics used throughout our simulation, including packet lengths, production intervals, and start time distributions for different application types:
\begin{table*}[]
    \centering
    \caption{Characterization of Application Types}
    \label{tab:app_characterization}
    \begin{tabular}{lccc}
        \hline
        \textbf{App Type} & \textbf{Packet Length} & \textbf{Production Interval} & \textbf{Start Time} \\ 
        \hline
        \multirow{2}{*}{\textbf{Network Control}} 
            & 498B & 55ms & uniform(0s, 0.1s) \\
            &      & Initial Offset: uniform(0ms, 5ms) & \\ 
        \hline
        \multirow{2}{*}{\textbf{Video}} 
            & 1453B & uniform(60ms, 65ms) & uniform(0.2s, 0.5s) \\
            &       & Initial Offset: uniform(0ms, 20ms) & \\
        \hline
        \multirow{2}{*}{\textbf{Best Effort}} 
            & 1429B & exponential(600ms) & uniform(0.5s, 1.0s) \\
            &       & Initial Offset: uniform(0ms, 100ms) & \\
        \hline
    \end{tabular}
\end{table*}
For the CBS, based on Table \ref{tab:app_characterization}, the video packet length is specified as 1453B, which is calculated as follows:

\begin{multline}
\text{Packet Length} = 1453B \ (\text{APP}) \ + 8B \ (\text{UDP}) \\
+ 20B \ (\text{IP}) \ + 14B \ (\text{ETH MAC}) \\
+ 4B \ (\text{ETH FCS}) \ + 8B \ (\text{ETH PHY})
\end{multline}

\begin{equation}
\text{Channel Packet Length} = 1507B
\end{equation}

\noindent The corresponding channel data rate can be calculated as

\begin{equation}
\text{Data Rate} = \frac{(1507B + 12B \, (\text{IFG}) + 1B \, (\text{cushion})) \times 8}{\text{Packet Interval}}
\end{equation}

\begin{figure}
    \centering
    \includegraphics[width=1\linewidth]{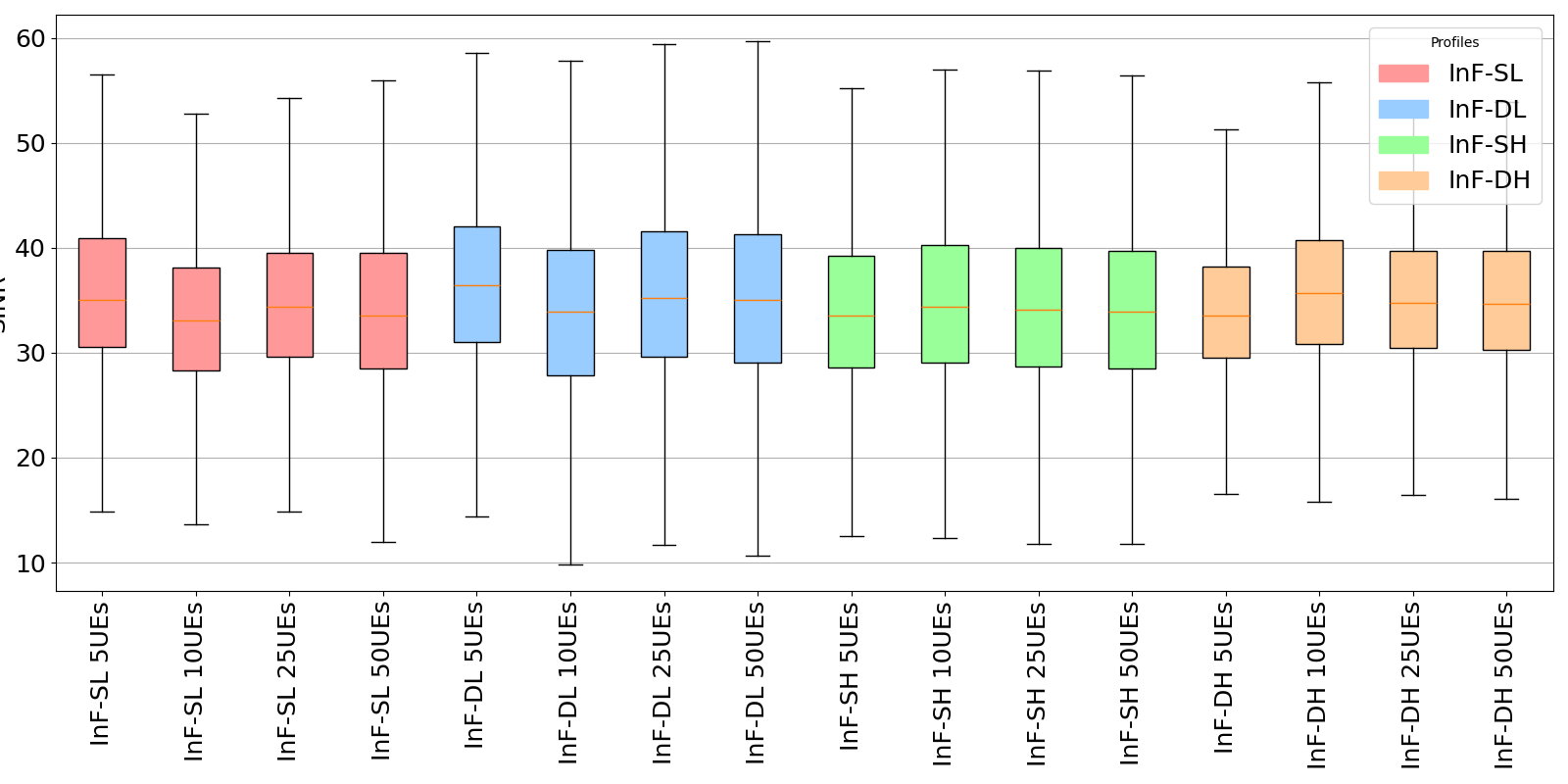}
    \caption{SINR Distribution by Profile and No. of UEs in downlink transmission}
    \label{fig:sinr-d2-downlink}
\end{figure}

\begin{figure}
    \centering
    \includegraphics[width=1\linewidth]{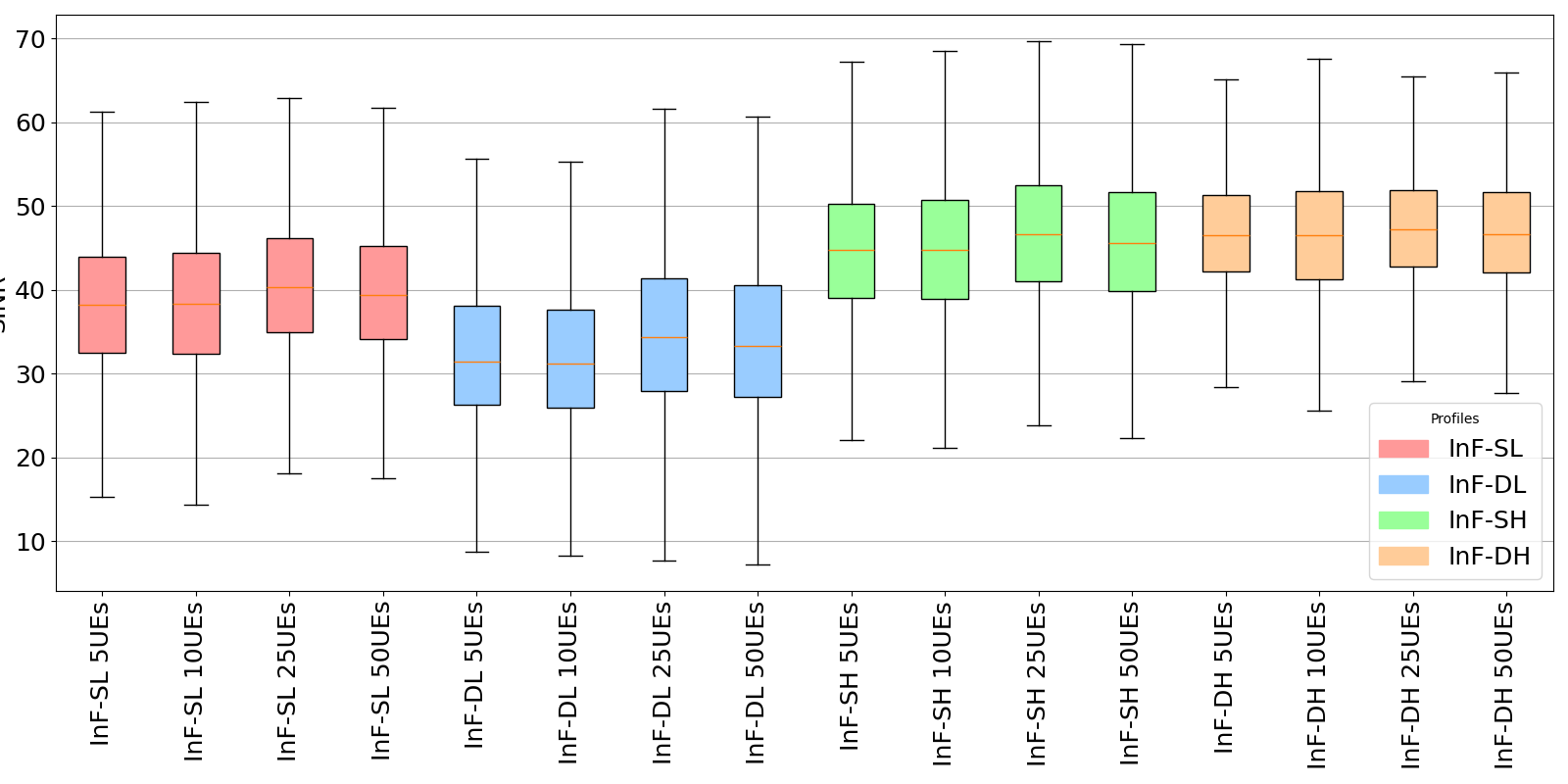}
    \caption{SINR Distribution by Profile and No. of UEs in uplink transmission}
    \label{fig:sinr-d2-uplink}
\end{figure}

Figs. \ref{fig:sinr-d2-downlink}, \ref{fig:sinr-d2-uplink} show the Signal Interference and Noise Ratio (SINR) of different InF profiles in the downlink and uplink transmission, respectively, each using its default parameter values.  
Based on Fig. \ref{fig:sinr-d2-downlink}, the InF-SL profile achieves the highest and most consistent SINR across all UE densities, demonstrating the advantage of low base station placement in sparse clutter environments. In contrast, the InF-DH profile consistently exhibits the lowest SINR values, underscoring the combined challenges posed by dense clutter and high base station placement. Furthermore, the variability in SINR increases with the number of UEs across all profiles, reflecting the impact of greater interference and resource contention as UE density grows.
As illustrated in Fig. \ref{fig:sinr-d2-uplink}, the uplink SINR results highlight that low base station placement (InF-SL) delivers the most reliable uplink performance, particularly in sparse clutter environments. In contrast, dense clutter and high base stations (InF-DH) lead to significant SINR degradation, especially as UE density increases.

\begin{figure}
    \centering
    \includegraphics[width=1\linewidth]{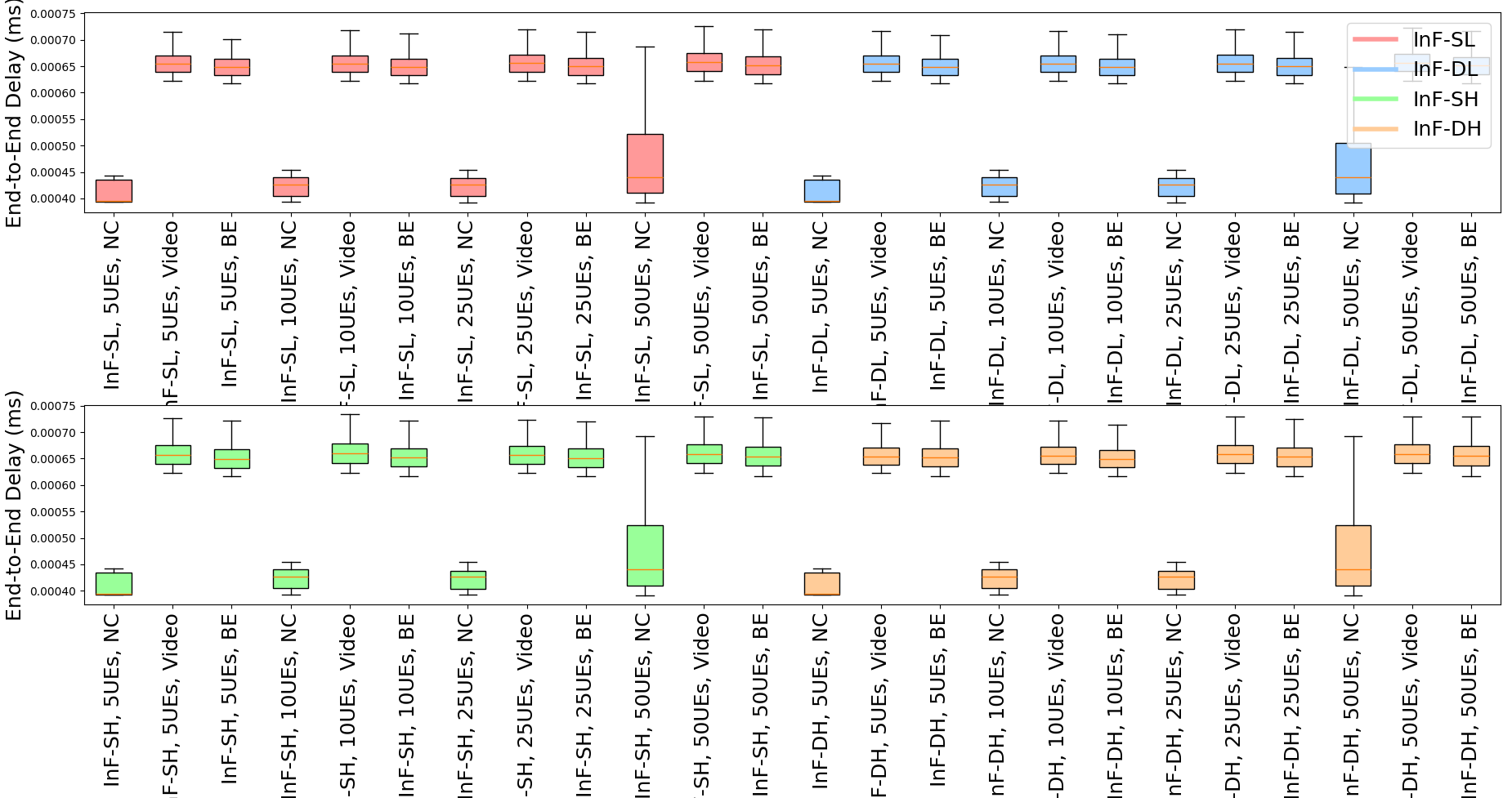}
    \caption{End-to-End delay for the downlink transmission across different InF profiles and varying numbers of UEs}
    \label{fig:E2E-d2-downlink}
\end{figure}

\begin{figure}
    \centering
    \includegraphics[width=1\linewidth]{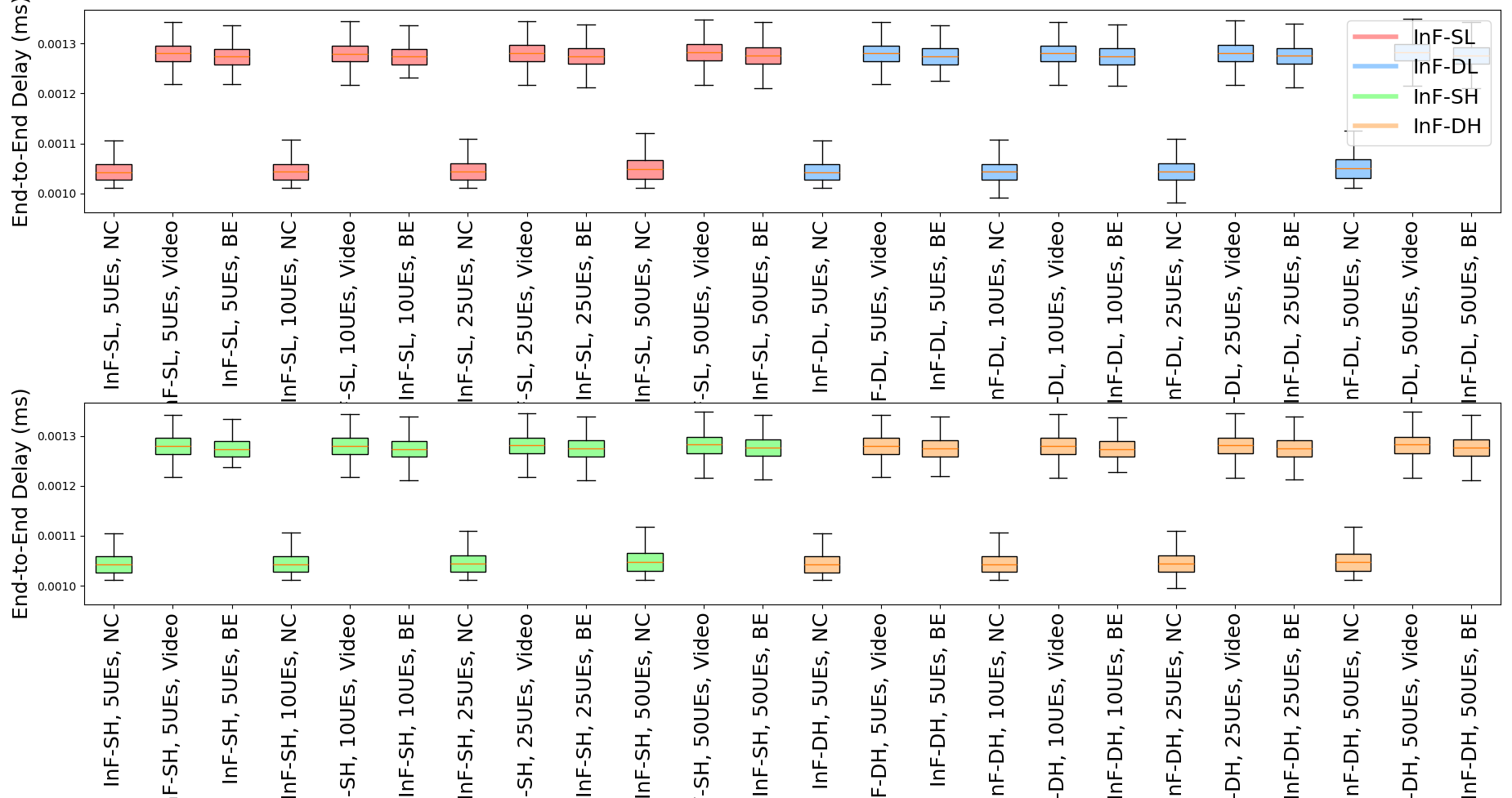}
    \caption{End-to-End delay for the uplink transmission across different InF profiles and varying numbers of UEs}
    \label{fig:E2E-d2-uplink}
\end{figure}

Figs. \ref{fig:E2E-d2-downlink}, \ref{fig:E2E-d2-uplink} show the influence of various InF profiles with different numbers of UEs on the end-to-end latency for each data stream for downlink and uplink transmission. The analysis considers node speeds ranging from $0.2$ m/s to $1.5$ m/s, utilizing Random Waypoint Mobility and a maximum distance (d2) constraint of 170 meters from the gNB. The end-to-end latency represents the time required for a data packet to travel from the source application to the destination application.
In Fig. \ref{fig:E2E-d2-downlink} the higher delay for NC at 50 UEs can be attributed to the increased contention for transmission resources. Specifically, with a higher number of UEs, the video traffic, which has larger packet sizes and relatively frequent production intervals, occupies significant transmission resources. As a result, NC packets, despite their low-latency requirements, are forced to wait for the ongoing video transmissions to complete. To mitigate the increased end-to-end delay for NC traffic at 50 UEs, a more optimized priority scheduling mechanism combined with preemptive scheduling can be employed. Optimized priority scheduling ensures that NC traffic is consistently prioritized over other traffic classes, such as Video and Best Effort, based on its stringent latency requirements. By integrating preemptive scheduling, NC packets can interrupt ongoing transmissions, particularly large Video packets, to meet critical latency deadlines. This combination reduces scheduling delays caused by resource contention and ensures that NC traffic experiences minimal waiting times, even under high UE densities. Such an approach can effectively improve network responsiveness and maintain the reliability of time-critical NC transmissions in scenarios with heavy traffic loads.

\begin{figure}
    \centering
    \includegraphics[width=1\linewidth]{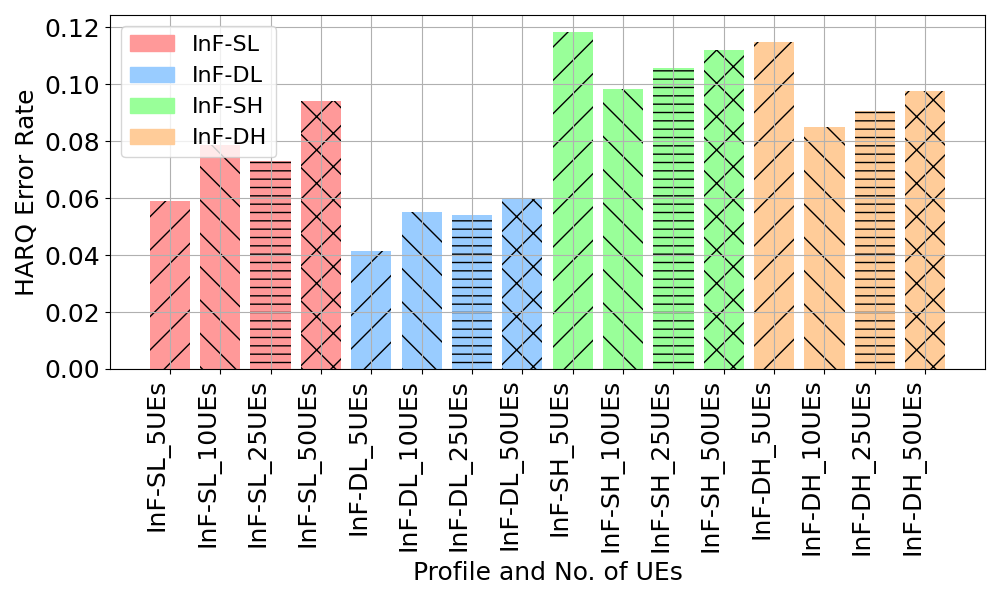}
    \caption{HARQ error rate with different profiles and No. of UEs for downlink}
    \label{fig:harqerror_d2_downlink}
\end{figure}

\begin{figure}
    \centering
    \includegraphics[width=1\linewidth]{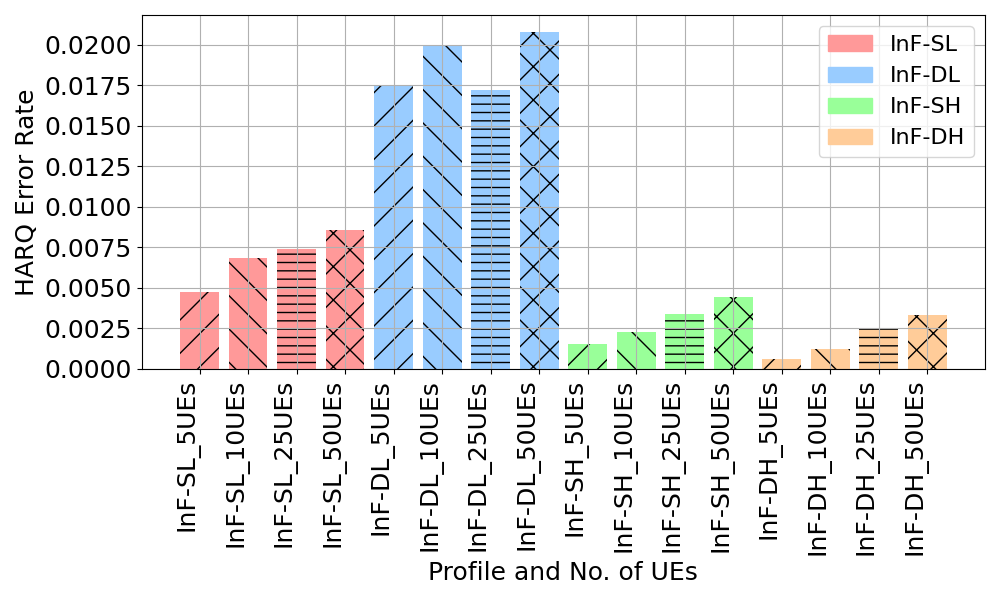}
    \caption{HARQ error rate with different profiles and No. of UEs for uplink}
    \label{fig:harqerror_d2_uplink}
\end{figure}

Fig. \ref{fig:harqerror_d2_downlink} presents the results for the downlink scenario in an indoor factory environment, using different profiles (InF-SL, InF-DL, InF-SH, and InF-DH) and varying numbers of UEs in each case, with the distance defined as d2 across all configurations. Fig. \ref{fig:harqerror_d2_uplink} illustrates the corresponding results for the uplink scenario.
These figures analyze the behaviour of the radio interface's Hybrid Automatic Repeat reQuest (HARQ) mechanism under varying indoor factory conditions. The profiles explore the effects of sparse/dense clutter and low/high base station placement on transmission reliability. The HARQ error rate, measured as the ratio of failed transmissions to the total number of transmissions, serves as an indicator of reliability for time-critical and non-critical data transmissions. This comparison highlights how varying clutter, base station heights, and UE densities influence error performance under the given distance conditions.

The HARQ error rate results for the downlink transmission in an indoor factory environment reveal distinct trends across various profiles and UE densities. The InF-SL profile exhibits moderate error rates, indicating a gradual decline in reliability as the number of UEs grows. In contrast, the InF-DL profile maintains lower error rates, demonstrating better resilience to increasing UE densities despite the presence of dense clutter. The InF-SH profile consistently shows the highest error rates, highlighting the impact of path loss due to the high base station placement. Similarly, the InF-DH profile experiences significant error rates, reflecting the combined challenges of multipath fading and obstruction caused by dense clutter. Overall, the results indicate that low base station placement (InF-SL and InF-DL) provides more reliable downlink performance, while high base station placement (InF-SH and InF-DH) leads to increased error rates, particularly in sparse clutter environments.


\section{Conclusion} \label{sec:Conclusion}
The transition to Industry 4.0 requires factories to replace traditional industrial networks, which primarily rely on Time Sensitive Ethernet (TSN) and other wired technologies, with wireless alternatives. This shift is critical for achieving the flexibility and adaptability demanded by modern manufacturing environments. While some leading companies have begun experimenting with 5G wireless communication, deployments remain in their infancy. In this study, we simulated a 3GPP compliant 5G Time Sensitive Network deployment model within an indoor factory setting to evaluate its feasibility for industrial applications. Simulations were conducted with various factory environment models to analyze their performance under different wireless channel conditions. Our findings demonstrate that 5G-TSN can reliably support latency-sensitive applications in controlled indoor environments, highlighting its potential for significant performance gains.

As our industrial indoor environment context is a novel exploration, our research establishes a baseline for evaluating 5G-TSN in such industrial scenarios. Future work will involve testing with more realistic conditions, incorporating variables such as increased gNBs, and factors like heavy network loads and dynamic environmental interference. These additional tests aim to evaluate the scalability, reliability, and robustness of 5G-TSN in more complex and dynamic factory settings. Furthermore, future efforts could focus on real-world deployment and validation to confirm simulation results and uncover practical challenges. Additional research may also investigate the integration of 5G-TSN with existing factory automation systems to ensure compatibility and fully exploit its advantages.

\section*{Acknowledgments}
This publication has emanated from research conducted with the financial support of Taighde Éireann – Research Ireland under Grant number 13/RC/2077\_P2 (CONNECT: the Research Ireland Centre for Future Networks).

\bibliography{references}
\bibliographystyle{IEEEtran}

\end{document}